\documentclass[aps,pre,twocolumn,nofootinbib,floatfix]{revtex4}
\usepackage{amsmath}
\usepackage{epsfig}

\input{tcilatex}

\begin{document}

\title{Current detection of superradiance and induced entanglement of double
quantum dot excitons}
\author{Y. N. Chen$^{1}$, D. S. Chuu$^{1}$, and T. Brandes$^{2}$}
\affiliation{$^{1}$Department of Electrophysics, National Chiao-Tung University, Hsinchu
30050, Taiwan\\
$^{2}$Department of Physics, UMIST, P.O. Box 88, Manchester, M60 1QD, U.K.}
\date{\today }

\begin{abstract}
We propose to measure the superradiance effect by observing the current
through a semiconductor double-dot ststem. An electron and a hole are
injected separately into one of the quantum dots to form an exciton and then
recombine radiatively. We find that the stationary current shows oscillatory
behavior as one varies the inter-dot distance. The amplitude of oscillation
can be increased by incorporating the system into a microcavity.
Furthermore, the current is suppressed if the dot distance is small compared
to the wavelength of the emitted photon. This photon trapping phenomenon
generates the entangled state and may be used to control the emission of
single photons at predetermined times.

\end{abstract}

\maketitle

\address{$^{1}$Department of Electrophysics, National Chiao Tung University,
Hsinchu 30050, Taiwan\\
$^{2}$Department of Physics, UMIST, P.O. Box 88, Manchester, M60 1QD, U.K.}

\address{$^{1}$Department of Electrophysics, National Chiao Tung University,
Hsinchu 30050, Taiwan} 
\address{$^{2}$Department of Physics, UMIST, P.O. Box 88, Manchester, M60
1QD, U.K.}

\address{Department of Electrophysics, National Chiao Tung University,
Hsinchu 300, Taiwan}

%\draft
%\wideabs{

% version ZI

%}

%%%%%%%%%%%%%%%%%%%%%%%%%%%%%%%%%%%%%%%%%%%%%%%%%%%%%%%%%%

Since Dicke proposed the phenomenon of superradiance\cite{1}, the coherent
radiation phenomena for atomic systems were intensively investigated. In
semiconductor systems, the electron-hole pair is naturally a candidate for
examining the spontaneous emission. The decay rate of the exciton is
superradiantly enhanced by a factor of $(\lambda /d)^{2}$ for a 2D
exciton-polariton system\cite{2}, where $\lambda $ is the wave length of the
emitted photon and $d$ is the lattice constant of the thin film. In the past
decades, the superradiance of excitons in these quantum well structures has
been investigated intensively\cite{3}. For lower dimensional systems, the
decay rate of the exciton is enhanced by a factor of $\lambda /d$ in a
quantum wire\cite{4}. In the quantum dots, although the decay rate is shown
to be proportional to R$^{2.1}$\cite{5} which confirms the theoretical
prediction\cite{6}, acceptable experimental data on quantum dot
superradiance are still not plentiful owing to the difficulty of techniques
in observing the enhanced spontaneous emission optically.

In recent years, great attention has been focused on the entanglement issue
since generation of highly entangled states is one of the fundamental
requirements for quantum information processing\cite{7}. Lots of works have
been associated with quantum-optic and atomic systems\cite{8}. However, due
to the scalability of the quantum processor, solid-state realizations of
such phenomena are the favored choices\cite{9}. In fact, the superradiance
effect can also generate entanglement by spontaneous emission. Therefore, a
different way is proposed to observe the superradiant effect and generate
the entangled states in this work. By injecting electrons and holes into one
of the quantum dots of a double dot system and controlling the gate voltage
of one of the dots, a photon is then generated by the coherent recombination
of the exciton. This process not only allows one to determine the
superradiant effect by measuring the stationary current, but also induces
the entangled states in this double-dot system.

In our model, we consider a p-i-n junction, similar to the device proposed
by Benson \textit{et al} \cite{10}, but with \emph{two spatially separated}
quantum dots. The novel feature here is the dissipative creation of
entanglement over relatively large distances, and its readout via the
stationary current The device structure is shown in Fig.1. 
\begin{figure}[th]
\includegraphics[width=8cm]{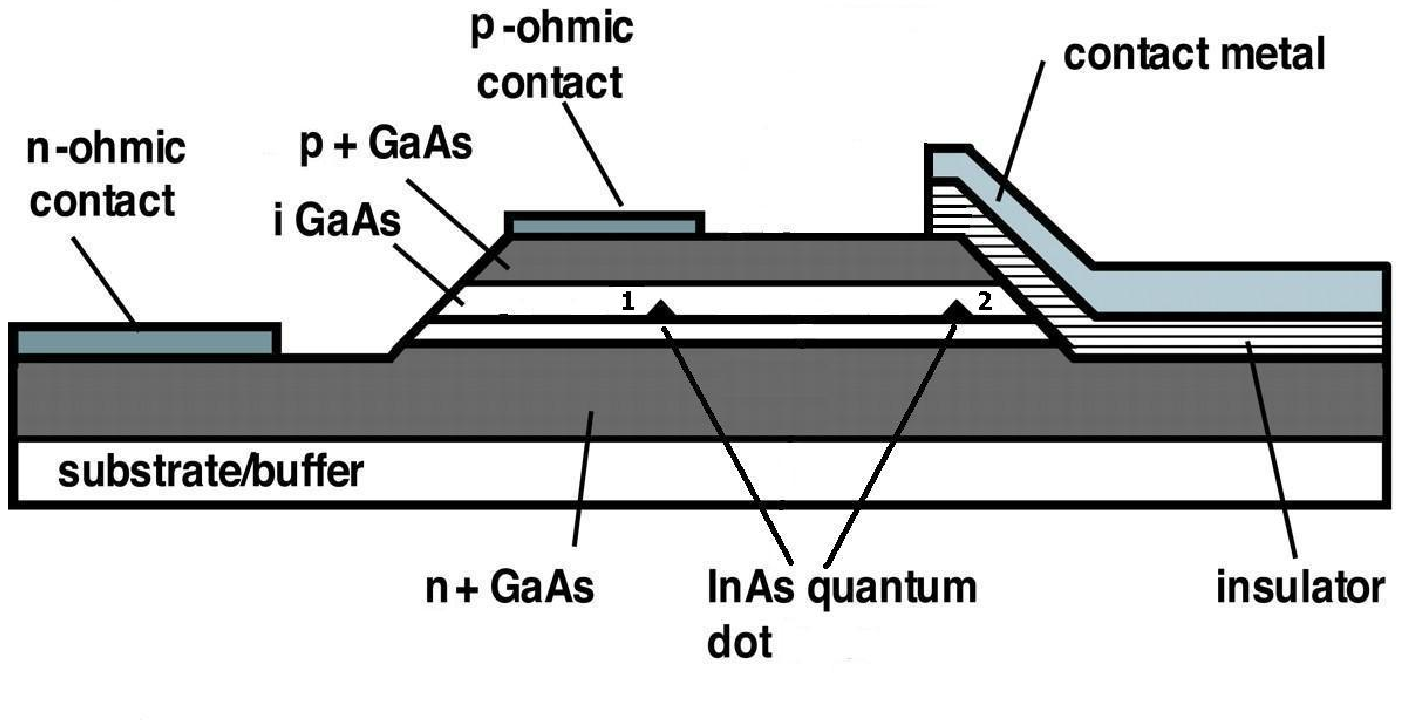}
\caption{Proposed device structure. Two InAs quantum dots are embedded in a%
\textit{\ p-i-n }junction. Above dot 2 is a metal gate, which control the
energy gap and orientation of the dipole.}
\end{figure}

One of the obstacles in measuring superradiance between the quantum dots
comes from the random size of the dots which result in a random distribution
of energy gap and thus diminishes the coherent radiation. This can be
overcome by constructing a gate voltage over one of the quantum dots. The
energy gap and the orientation of the dipole moments of one of the quantum
dots, thus can be controlled well. Both the hole and electron reservoirs are
assumed to be in thermal equilibrium. For the physical phenomena we are
interested in, the current is conducted through dot 1 only, and the fermi
level of the \textit{p(n)}-side hole (electron) is slightly lower (higher)
than the hole (electron) subband in the dot. After a hole is injected into
the hole subband in the quantum dot, the \textit{n}-side electron can tunnel
into the exciton level because of the Coulomb interaction between the
electron and hole. In our calculation, we also neglect the Forster process
which may have some influences on the results if the two dots are close to
each other. The validity of this assumption will be discussed later. Thus,
we may assume four dot states $\left| 0\right\rangle =\left|
0,h;0,0\right\rangle $, $\left| U_{1}\right\rangle =\left|
e,h;0,0\right\rangle $, $\left| U_{2}\right\rangle =\left|
0,0;e,h\right\rangle $, and $\left| D\right\rangle =\left|
0,0;0,0\right\rangle $, where $\left| 0,h;0,0\right\rangle $ means there is
one hole in dot 1 and $\left| 0,0;0,0\right\rangle $ represents the ground
state with no hole and electron in the quantum dots. The exciton states $%
\left| e,h;0,0\right\rangle $ (in dot 1) can be converted to $\left|
0,0;e,h\right\rangle $ (in dot 2) through the exciton-photon interactions.
One might argue that one can not neglect the state $\left|
e,0;0,0\right\rangle $ for real device since the tunable variable is the
applied voltage. However, this can be resolved by fabricating a thicker
barrier on the electron side so that there is little chance for an electron
to tunnel in advance. By transforming $\left| U_{1}\right\rangle $ and $%
\left| U_{2}\right\rangle $ into Dicke states: $\left| S_{0}\right\rangle =%
\frac{1}{\sqrt{2}}\left( \left| U_{1}\right\rangle -\left|
U_{2}\right\rangle \right) $ and $\left| T_{0}\right\rangle =\frac{1}{\sqrt{2%
}}\left( \left| U_{1}\right\rangle +\left| U_{2}\right\rangle \right) $, we
can now define the dot-operators $\overset{\wedge }{n_{S}}\equiv \left|
S_{0}\right\rangle \left\langle S_{0}\right| ,\overset{\wedge }{n_{T}}\equiv
\left| T_{0}\right\rangle \left\langle T_{0}\right| ,$ $\overset{\wedge }{%
n_{D}}\equiv \left| D\right\rangle \left\langle D\right| ,$ $\overset{\wedge 
}{p_{s}}\equiv \left| S_{0}\right\rangle \left\langle D\right| ,$ $\overset{%
\wedge }{p_{T}}\equiv \left| T_{0}\right\rangle \left\langle D\right| ,%
\overset{\wedge }{s_{U_{1}}}\equiv \frac{1}{\sqrt{2}}(\left| 0\right\rangle
\left\langle S_{0}\right| +\left| 0\right\rangle \left\langle T_{0}\right|
), $ $\overset{\wedge }{s_{D}}\equiv \left| 0\right\rangle \left\langle
D\right| $. The total hamiltonian $H$ of the system consists of three parts:$%
H_{0}$ (dot, photon bath $H_{p}$, electron (hole) reservoirs $H_{res}$), $%
H_{T}$ (dot-photon coupling), and the dot-reservoir coupling $H_{V}$:

\begin{eqnarray}
H &=&H_{0}+H_{T}+H_{V}  \notag \\
H_{0} &=&\varepsilon _{U}\overset{\wedge }{n_{S}}+\varepsilon _{U}\overset{%
\wedge }{n_{T}}+\varepsilon _{D}\overset{\wedge }{n_{D}}+H_{p}+H_{res} 
\notag \\
H_{T} &=&\sum_{\mathbf{k}}\frac{1}{\sqrt{2}}g\{D_{\mathbf{k}}b_{\mathbf{k}}[%
\overset{\wedge }{p_{S}}(1+e^{i\mathbf{k}\cdot \mathbf{r}})  \notag \\
&&+\overset{\wedge }{p_{T}}(1-e^{i\mathbf{k}\cdot \mathbf{r}})]+c.c.\} 
\notag \\
&=&g(\overset{\wedge }{p_{S}}X_{S}+\overset{\wedge }{p_{S}}^{\dagger
}X_{S}^{\dagger }+\overset{\wedge }{p_{T}}\overline{X_{T}}+\overset{\wedge }{%
p_{T}}^{\dagger }\overline{X_{T}}^{\dagger })  \notag \\
H_{p} &=&\sum_{\mathbf{k}}\omega _{\mathbf{k}}b_{\mathbf{k}}^{\dagger }b_{%
\mathbf{k}}  \notag \\
H_{V} &=&\sum_{\mathbf{q}}(V_{\mathbf{q}}c_{\mathbf{q}}^{\dagger }\overset{%
\wedge }{s_{U_{1}}}+W_{\mathbf{q}}d_{\mathbf{q}}^{\dagger }\overset{\wedge }{%
s_{D}}+c.c.)  \notag \\
H_{res} &=&\sum_{\mathbf{q}}\varepsilon _{\mathbf{q}}^{U}c_{\mathbf{q}%
}^{\dagger }c_{\mathbf{q}}+\sum_{\mathbf{q}}\varepsilon _{\mathbf{q}}^{D}d_{%
\mathbf{q}}^{\dagger }d_{\mathbf{q}}.
\end{eqnarray}%
In the above equation, $b_{\mathbf{k}}$ is the photon operator, $gD_{\mathbf{%
k}}$ is the dipole coupling strength, $\mathbf{r}$ is the position vector
between two quantum dot, $X_{S}=\sum_{\mathbf{k}}(1+e^{i\mathbf{k}\cdot 
\mathbf{r}})D_{\mathbf{k}}b_{\mathbf{k}}$ , $\overline{X_{T}}=\sum_{\mathbf{k%
}}(1-e^{i\mathbf{k}\cdot \mathbf{r}})D_{\mathbf{k}}b_{\mathbf{k}}$,$\ $and $%
c_{\mathbf{q}}$ and $d_{\mathbf{q}}$ denote the electron operators in the
left and right reservoirs, respectively. The dipole approximation is not
used in our calculation since we keep the full $e^{i\mathbf{k}\cdot \mathbf{r%
}}$ terms in the operators $X_{S}$ and $\overline{X_{T}}$. Here, $g$ is a
constant with a unit of the tunneling rate. The couplings to the electron
and hole reservoirs are given by the standard tunnel hamiltonian $H_{V},$
where $V_{\mathbf{q}}$ and $W_{\mathbf{q}}$ couple the channels $\mathbf{q}$
of the electron and the hole reservoirs. If the couplings to the electron
and the hole reservoirs are weak, then it is reasonable to assume that the
standard Born-Markov approximation with respect to these couplings is valid.
In this case, one can derive a master equation from the exact time-evolution
of the system. The equations of motion can be expressed as (cp. [23])

\begin{eqnarray}
\overset{\wedge }{\left\langle n_{\sigma }\right\rangle }_{t}-\overset{%
\wedge }{\left\langle n_{\sigma }\right\rangle }_{0}
&=&-ig\int_{0}^{t}dt^{\prime }\{\overset{\wedge }{\left\langle p_{\sigma
}\right\rangle }_{t^{\prime }}-\overset{\wedge }{\left\langle p_{\sigma
}^{\dagger }\right\rangle }_{t^{\prime }}\}  \notag \\
&&+\Gamma _{U}\int_{0}^{t}dt^{\prime }(1-\overset{\wedge }{\left\langle
n_{S}\right\rangle }_{t^{\prime }}-\overset{\wedge }{\left\langle
n_{T}\right\rangle }_{t^{\prime }}-\overset{\wedge }{\left\langle
n_{D}\right\rangle }_{t^{\prime }})  \notag \\
\overset{\wedge }{\left\langle n_{D}\right\rangle }_{t}-\overset{\wedge }{%
\left\langle n_{D}\right\rangle }_{0} &=&-ig\int_{0}^{t}dt^{\prime }\{%
\overset{\wedge }{\left\langle p_{S}\right\rangle }_{t^{\prime }}-\overset{%
\wedge }{\left\langle p_{S}^{\dagger }\right\rangle }_{t^{\prime }}+\overset{%
\wedge }{\left\langle p_{T}\right\rangle }_{t^{\prime }}  \notag \\
&&-\overset{\wedge }{\left\langle p_{T}^{\dagger }\right\rangle }_{t^{\prime
}}\}-2\Gamma _{D}\int_{0}^{t}dt^{\prime }\overset{\wedge }{\left\langle
n_{D}\right\rangle }_{t^{\prime }}  \notag \\
\overset{\wedge }{\left\langle p_{S}\right\rangle }_{t}-\overset{\wedge }{%
\left\langle p_{S}\right\rangle _{t}^{0}} &=&-\Gamma
_{D}\int_{0}^{t}dt^{\prime }e^{i\varepsilon (t-t^{\prime })}\left\langle
X_{t}X_{t^{\prime }}^{\dagger }\widetilde{p_{S}}(t^{\prime })\right\rangle
_{t^{\prime }}  \notag \\
&&-ig\int_{0}^{t}dt^{\prime }e^{i\varepsilon (t-t^{\prime })}\{\left\langle 
\overset{\wedge }{n_{S}}X_{t}X_{t^{\prime }}^{\dagger }\right\rangle
_{t^{\prime }}  \notag \\
&&-\left\langle \overset{\wedge }{n_{D}}X_{t^{\prime }}^{\dagger
}X_{t}\right\rangle _{t^{\prime }}\}  \notag \\
\overset{\wedge }{\left\langle p_{T}\right\rangle }_{t}-\overset{\wedge }{%
\left\langle p_{T}\right\rangle _{t}^{0}} &=&-\Gamma
_{D}\int_{0}^{t}dt^{\prime }e^{i\varepsilon (t-t^{\prime })}\left\langle 
\overline{X}_{t}\overline{X}_{t^{\prime }}^{\dagger }\widetilde{p_{T}}%
(t^{\prime })\right\rangle _{t^{\prime }}  \notag \\
&&-ig\int_{0}^{t}dt^{\prime }e^{i\varepsilon (t-t^{\prime })}\{\left\langle 
\overset{\wedge }{n_{T}}\overline{X}_{t}\overline{X}_{t^{\prime }}^{\dagger
}\right\rangle _{t^{\prime }}  \notag \\
&&-\left\langle \overset{\wedge }{n_{D}}\overline{X}_{t^{\prime }}^{\dagger }%
\overline{X}_{t}\right\rangle _{t^{\prime }}\}
\end{eqnarray}%
, where the index $\sigma =S$ or $T$, $\Gamma _{U}$ $=\pi \sum_{\mathbf{q}%
}V_{\mathbf{q}}^{2}\delta (\varepsilon _{U}-\varepsilon _{\mathbf{q}}^{U})$
, $\Gamma _{D}=\pi \sum_{\mathbf{q}}W_{\mathbf{q}}^{2}\delta (\varepsilon
_{D}-\varepsilon _{\mathbf{q}}^{D})$, and $\varepsilon =\varepsilon
_{U}-\varepsilon _{D}$ is the energy gap of the quantum dot exciton. Here, $%
\widetilde{p_{S}}(t^{\prime })=p_{S}e^{i\varepsilon t}X_{t^{\prime }}$, $%
\widetilde{p_{T}}(t^{\prime })=p_{T}e^{i\varepsilon t}\overline{X}%
_{t^{\prime }}$, and $X_{t^{\prime }}$ ($\overline{X}_{t^{\prime }}$)
denotes the time evolution of $X$ \ ($\overline{X}$) with $H_{p}$. The
expectation value $\overset{\wedge }{\left\langle p_{\sigma }^{\dagger
}\right\rangle _{t}^{0}}$ describes the decay of an initial polarization of
the system and plays no role for the stationary current. Therefore, we shall
assume the initial expectation value of $\overset{\wedge }{p}_{\sigma
}^{(\dagger )}$ vanishes at time $t=0$.

As can be seen from Eq. (2), there are terms like $\left\langle \overset{%
\wedge }{n_{\sigma }}X_{t}X_{t^{\prime }}^{\dagger }\right\rangle
_{t^{\prime }}$ which contain products of dot operators and photon
operators. If we are interested in small coupling parameters here, a
decoupling of the reduced density matrix $\widetilde{\rho _{\sigma }}%
(t^{\prime })$ can be used, $\widetilde{\rho _{\sigma }}(t^{\prime })\approx
\rho _{ph}^{0}Tr_{ph}\widetilde{\rho _{\sigma }}(t^{\prime })$. Products of
these operators can then be obtained. For spontaneous emission, the photon
bath is assumed to be in equilibrium. The expectation values $\left\langle
X_{t}X_{t^{\prime }}^{\dagger }\right\rangle _{0}\equiv C(t-t^{\prime })$
and $\left\langle \overline{X}_{t}\overline{X}_{t^{\prime }}^{\dagger
}\right\rangle _{0}\equiv \overline{C}(t-t^{\prime })$ are functions of the
time difference only. We can now define the Laplace transformation for real $%
z,$

\begin{eqnarray}
C_{\varepsilon }(z) &\equiv &\int_{0}^{\infty }dte^{-zt}e^{i\varepsilon
t}C(t)  \notag \\
n_{\sigma }(z) &\equiv &\int_{0}^{\infty }dte^{-zt}\overset{\wedge }{%
\left\langle n_{\sigma }\right\rangle }_{t}\text{ \ }etc.,\text{ }z>0
\end{eqnarray}%
and transform the whole equations of motion into $z$-space. The tunnel
current $\widehat{I}$ can be defined as the change of the occupation of $%
\overset{\wedge }{n_{D}}$ and is given by $\widehat{I}\equiv ig\sum_{\sigma
}(\overset{\wedge }{p_{\sigma }}-\overset{\wedge }{p_{\sigma }}^{\dagger }),$
where we have set the electron charge $e=1$ for convenience. The time
dependence of the expectation value $\overset{\wedge }{\left\langle
I\right\rangle }_{t}$ can be obtained by performing the inverse Laplace
transformation. For time $t\rightarrow \infty ,$ the result is

\begin{equation}
\overset{\wedge }{\left\langle I\right\rangle }_{t\rightarrow \infty }=\frac{%
4g^{2}\gamma _{+}\gamma _{-}}{\gamma _{-}+\gamma _{+}[1+2\gamma
_{-}(g^{2}/\Gamma _{D}+g^{2}/\Gamma _{U}+\Gamma _{D})]}
\end{equation}%
, where $g^{2}\gamma _{+}$ and $g^{2}\gamma _{-}$ are the superradiant and
subradiant decay rate of the exciton, respectively.\cite{11} The derivation
of the current equation is closely analogous to the spontaneous emission of
phonons in double dots\cite{12}, in which the correlation functions $%
\left\langle X_{t}X_{t^{\prime }}^{\dagger }\right\rangle _{0}$ is given by
the electron-phonon interaction.

The corresponding decay rate for superradiant and the subradiant channels is
given by

\begin{equation}
g^{2}\gamma _{\pm }=\gamma _{0}(1\pm \frac{\sin (2\pi d/\lambda _{0})}{2\pi
d/\lambda _{0}}),
\end{equation}%
where $d$ is the inter-dot distance and $\gamma _{0}$ is the exciton decay
rate in a quantum dot. To display the dependence of the stationary current
through the quantum dot on the dot distance $d$, we present the results of
two identical quantum dots in Fig. 2. 
\begin{figure}[h]
\includegraphics[width=7.5cm]{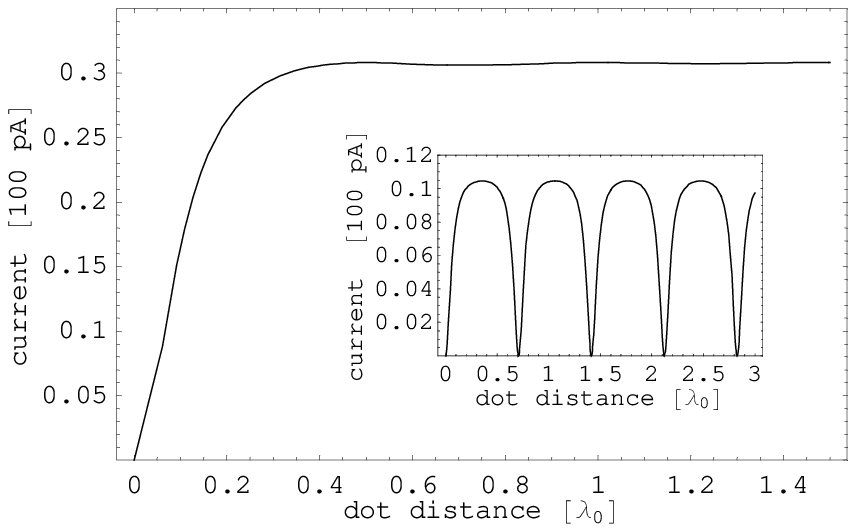}
\caption{{}Stationary tunnel current, Eq. (4), as a function of dot distance 
$d$. The interference effect is seen clearly (inset) by incorporating the
system inside a rectangular microcavity. The vertical and horizontal units
are 100 pA and $\protect\lambda _{0}$, respectively.}
\end{figure}
In plotting the figure, the current is in terms of 100 pA. Furthermore, the
tunneling rates, $\Gamma _{U}$ and $\Gamma _{D}$, are assumed to be equal to 
$0.2\gamma _{0}$ and $\gamma _{0},$ respectively. Here, a value of 1/1.3ns
for the free-space quantum dot decay rate $\gamma _{0}$ is used in our
calculations\cite{13}. As shown in Fig. 2, the current is suppressed as the
dot distance $d$ is much smaller than the wavelength ($\lambda _{0}$) of the
emitted photon. This corresponds to the trapping state in the two-ion
system. As long as we choose only one of the dots to be coupled to
reservoirs, the generated photon is reabsorbed immediately by the other dot
and vice versa. The current is then blocked by this exchange process. For
small rates limit ($g^{2}\gamma _{\pm }$) one can approximate equation (4)
by $I\approx 4[1/g^{2}\gamma _{-}+1/g^{2}\gamma _{+}]^{-1}$. The rates $%
\Gamma _{U,D}$ drop out completely and the current is only determined by the
(smaller) radiative decay rates. In this approximate form, the current looks
identical to the expression for the conductance $G\propto \lbrack 1/\Gamma
_{L}+1/\Gamma _{R}]^{-1}$ from a left lead through a single level to a right
lead with tunnel rates $\Gamma _{L,R}$. This implies that the superradiant
and the subradiant channel are in series (and not in parallel) in this
limit. This is because once the exciton is formed in dot 1, time evolution
of this state is proportional to $e^{-g^{2}\gamma _{+}t}+e^{-g^{2}\gamma
_{-}t}$ not $e^{-g^{2}(\gamma _{+}+\gamma _{-})t}$\cite{14}. It means the
two decay channels in our system are not in parallel. For long time behavior 
$t\rightarrow \infty $ and $\gamma _{+}>>\gamma _{-}$, the function $%
e^{-g^{2}\gamma _{+}t}+e^{-g^{2}\gamma _{-}t}$ approaches the limit of $%
e^{-g^{2}\gamma _{-}t}$, which is identical to the same limit of the
function $e^{-\frac{g^{2}\gamma _{+}\gamma _{-}}{\gamma _{+}+\gamma _{-}}t}$%
(in series).

Similar to the two-ion superradiance\cite{11}, the current also exhibits
oscillatory behavior as a function of dot distance. To observe the
interference effect clearly, one may incorporate the system inside a
microcavity since semiconductor cavities with strong electron-photon
coupling have been realised experimentally by, e.g., G\'{e}rard\textit{\ et
al}.\cite{15}. Reduction of the allowed $k$-state is expected to increase
the magnitude of the oscillation. For example, if the system is placed
inside a rectangular microcavity with length $\lambda _{0}$, the decay rate
for the two channels can be worked out straightforwardly:

\begin{equation}
g^{2}\gamma _{cav,\pm }=\frac{\gamma _{0}}{\pi }\left| 1\pm e^{i2\pi d/(%
\sqrt{2}\lambda _{0})}\right| ^{2}.
\end{equation}%
The stationary current is plotted in the inset of Fig. 2, where a perfect
(lossless) cavity is assumed. As we mentioned above, the amplitude of
oscillation is larger than that in free space. However, the oscillation
period is not half of the wavelength, but $\lambda _{0}/\sqrt{2}$. This is
because the interference term is only influenced by the wave vector in the
unconfined direction. Excluding the contributions from fundamental cavity
modes, the effective wave vector can be expressed as

\begin{equation}
k_{eff}=\sqrt{(\frac{2\pi }{\lambda _{0}})^{2}-2\ast (\frac{\pi }{\lambda
_{0}})^{2}}=\frac{k_{0}}{\sqrt{2}}.
\end{equation}%
The oscillation period of the decay rate and the current is therefore
increased by a factor of $\sqrt{2}$.

In Fig. 3, we plot the expectation value of $n_{S\text{ }}$ ($n_{T}$) as a
function of the dot distance. The maximum entangled state ($\left|
S_{0}\right\rangle $) is reached as $d<<$ $\lambda _{0}$. This is remarkable
as the steady state is independent of the initial state. The entanglement is
induced by the cooperative decoherence in the system. In a recent paper by
Schnider \textit{et al}.\cite{16}, the authors consider the behavior of an
ion trap with all ions driven simultaneously and coupled collectively to a
heat bath. They also found that the steady state of the ion trap can exhibit
quantum entanglement. However, the concurrence of their system is below the
value of unity (maximum entanglement). On the contrary, in our system the
maximum entangled state can be generated by tuning the band gap of dot 2
(linear stark effect), i.e. control the on/off of the superradiance. Another
advantage of our scheme is shown in the inset of Fig. 3. If the double-dot
system is incorporated inside a rectangular microcavity, the maximum
entangled states repeat as a function of inter-dot distance. This means even
for remote separation, the entanglement can still be achieved. The reason
can be attributed to that the creation of entanglement in our model is
governed by the interaction with a common heat bath, while conventional
creation of entanglement depends on the direct interaction between two
subsystems\cite{17}. When two dots are coupled to the common photon fields,
the collective decay process drives the system into the entangled states.
The novel feature of the effect predicted here is that entanglement in fact
can be controlled electrically (without applying a laser field) and read out
in the form of a transport property, i.e., the electron \emph{current} (as a
function of the dot distance or, alternatively, the cavity length). 
\begin{figure}[h]
\includegraphics[width=7.5cm]{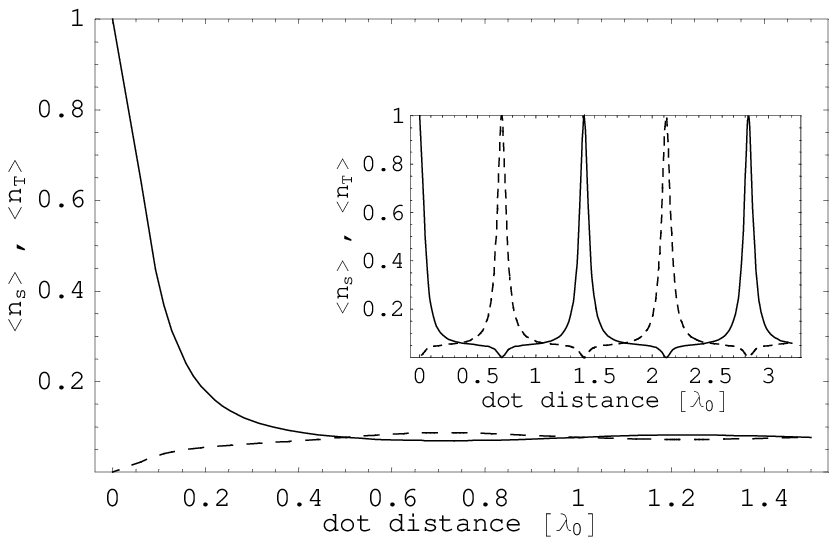}
\caption{{}Occupation probability of the entangled states $n_{S\text{ }}$
(solid line) and $n_{T}$ (dashed line). The inset shows the results inside a
rectangular microcavity.}
\end{figure}

Another possible application of this effect is that by tuning the coherence
of the dots, one can control the emission of single photon at predetermined
times, which is important for the field of quantum information technology.
However, one should note that the biexciton effect is not included in our
present model. Therefore, a low injection limit is required in the experiment%
\cite{18}. One might argue that for small inter-dot distance the Forster
process may play some role in our system\cite{19}; nevertheless, this only
causes small energy splitting between state $\left| S_{0}\right\rangle $ and 
$\left| T_{0}\right\rangle $. Comparing to the large energy difference in
the III-V semiconductor material, its effect on the decay rate $g^{2}\gamma
_{\pm }$ is negligible. As for the problem of dissipation, decoherence due
to interaction with other bosonic excitations (phonons and electron-hole
pairs in the leads) is inevitable but can in principle be (partly)
controlled by variation of the dot energies, or control of the mechanical
degree of freedom\cite{20}. In addition, scattering due to impurities are
negligible since there is no interdot transport in our system.

In conclusion, we have proposed a method of detecting the superradiant
effect in a semiconductor double-dot system. By incorporating the InAs
quantum dots between a \textit{p-i-n} junction, the superradiant effect on
the stationary tunnel current can be examined by tuning the band gap of the
quantum dot. Moreover, the interference effects between two dots can be seen
more clearly by incorporating the system inside a microcavity. The
oscillation period of the decay rate and current is also increased because
of the microcavity. Besides, the maximum entangled state is induced as the
inter-dot distance is much smaller than the wavelength of the emitted
photon. Our model provides a new way to generate the entanglement in
solid-state systems.

We would like to thank to Prof. Y. C. Lee at SUNYAB for helpful discussions.
This work is supported partially by the National Science Council, Taiwan
under the grant number NSC 91-2112-M-009-012.


\begin{thebibliography}{99}
\bibitem{1} R. H. Dicke, Phys. Rev. \textbf{93}, 99 (1954).

\bibitem{2} K. C. Liu and Y. C. Lee, Physica \textbf{102A}, 131 (1980); Y.
N. Chen and D. S. Chuu, Phys. Rev. \textbf{B 61}, 10815 (2000).

\bibitem{3} J. Knoester, Phys. Rev. Lett.\textbf{\ 68}, 654 (1992); D. S.
Citrin, Phys. Rev. B \textbf{47}, 3832 (1993); D. Ammerlahn, J. Kuhl, B.
Grote, S. W. Koch, G. Khitrova, and H. Gibbs, Phys. Rev. B \textbf{62}, 7350
(2000).

\bibitem{4} A. L. Ivanov and H. Haug, Phys. Rev. Lett.\textbf{\ 71}, 3182
(1993); Y. Manabe, T. Tokihiro, and E. Hanamura, Phys. Rev. \textbf{B 48},
2773 (1993); Y. N. Chen and D. S. Chuu, T. Brandes, and B. Kramer, Phys.
Rev. \textbf{B 64}, 125307 (2001).

\bibitem{5} A. Nakamura, H. Yamada, and T. Tookizaki, Phys. Rev. \textbf{B 40%
}, 8585 (1989).

\bibitem{6} E. Hanamura, Phys. Rev. \textbf{B 38}, 1228 (1988); F. C. Spano,
J. R. Kuklinski, and S. Mukamel, Phys. Rev. Lett. \textbf{65,} 211 (1990).

\bibitem{7} C. H. Bennett and D. P. DiVincenzo, Nature (London) \textbf{404}%
, 247 (2000).

\bibitem{8} T. Pellizzari, S. A. Gardiner, J. I. Cirac, and P. Zoller, Phys.
Rev. Lett. 75, 3788 (1995); J. I. Cirac and P. Zoller, Phys. Rev. Lett. 
\textbf{74}, 4091 (1995); K. Molmer and A. Sorensen, Phys. Rev. Lett. 
\textbf{82}, 1835 (1999).

\bibitem{9} A.T. Costa, Jr. and S. Bose, Phys. Rev. Lett. 87, 277901 (2001);
W.D. Oliver, F. Yamaguchi, and Y. Yamamoto, Phys. Rev. Lett. \textbf{88},
037901 (2002); Oliver Gywat, Guido Burkard, and Daniel Loss, Phys. Rev. 
\textbf{B 65}, 205329 (2002).

\bibitem{10} O. Benson, C. Santori, M. Pelton, and Y. Yamamoto, Phys. Rev.
Lett. \textbf{84}, 2513 (2000).

\bibitem{11} R. G. DeVoe and R. G. Brewer, Phys. Rev. Lett. \textbf{76},
2049 (1996).

\bibitem{12} T. Brandes and B. Kramer, Phys. Rev. Lett. \textbf{83}, 3021
(1999).

\bibitem{13} G. S. Solomon, M. Pelton, and Y. Yamamoto, Phys. Rev. Lett. 
\textbf{86}, 3903 (2001).

\bibitem{14} Paul S. Lee, Y. C. Lee, and C. T. Chang , Phys. Rev. \textbf{A} 
\textbf{8}, 1722 (1973).

\bibitem{15} J. M. G\'{e}rard, B. Sermage, B. Gayral, B.Legrand, E. Costard,
and V. Thierry-Mieg, Phys. Rev. Lett. \textbf{81}, 1110 (1998).

\bibitem{16} S. Schneider and G. J. Milburn, Phys. Rev. \textbf{A} \textbf{65%
}, 042107 (2002).

\bibitem{17} Daniel Braun, Phys. Rev. Lett. \textbf{89}, 277901 (2002).

\bibitem{18} Z. Yuan, B. E. Kardynal, R. M. Stevenson, A. J. Shields, C. J.
Lobo, K. Cooper, N. S. Beattie, D. A. Ritchie, M. Pepper1, Science \textbf{%
295}, 102 (2002).

\bibitem{19} L. Quiroga and N. Johnson, Phys. Rev. Lett. \textbf{83}, 2270
(1999).

\bibitem{20} S. Debald, T. Brandes, and B. Kramer, Phys. Rev. \textbf{B} 
\textbf{66}, 041301(\textbf{R}) (2002).
\end{thebibliography}
\end{document}